**Intercross: An Android app for plant breeding and genetics cross management**

Trevor W. Rife[1*+], Chaney Courtney[1+], Guillaume Bauchet[2,3], Mitchell Neilsen[4], and Jesse A. Poland[1]

[1] Wheat Genetics Resource Center, Dep. of Plant Pathology, 4024 Throckmorton Plant Sciences Center, 1712 Claflin Road, Kansas State Univ., Manhattan, KS 66506

[2] Boyce Thompson Institute, 533 Tower Road, Cornell University, Ithaca, NY 14853

[3] Current address: Terre de Lin, 605 Route de la Vallée, 76740 Saint-Pierre-le-Viger, France

[4] Dep. of Computer Science, 2184 Engineering Hall, 1701D Platt St., Kansas State University, Manhattan, KS 66506

[+] These authors contributed equally to this work

* Corresponding author:
T.W. Rife: trife@ksu.edu

# ABSTRACT

Plant breeding is fundamentally comprised of three cyclic activities: 1) intermating lines to generate novel allelic combinations, 2) evaluation of new plant cultivars in distinct environments, and 3) selection of superior individuals to be used as parents in the next breeding cycle. While digital technologies and tools are commonly utilized for the latter two stages, many plant research programs still rely on manual annotation and paper tags to track the crosses that constitute the basis of a plant breeding program. This presence of analog data is a crack in the foundation of a digital breeding ecosystem and a significant occasion for errors to be introduced that will propagate through the entire breeding program. However, implementing digital cross tracking into breeding programs is difficult due to the non-standardized workflows that different breeders have adopted. Intercross, an open-source Android app, aims to provide scientists with a robust and simple solution for planning, tracking, and managing the crosses being made each season and aims to serve as the primary tool to digitize crossing data for breeding programs. The simplicity and flexibility of Intercross allows rapid and broad adoption by diverse breeding programs and will solidify the concepts of a digital breeding ecosystem.

# INTRODUCTION

Digital data collection in plant breeding and genetics research is fundamentally necessary to achieve the throughput and accuracy required to identify and select improved varieties. Plant population improvement is ultimately a repeating cycle of evaluating experimental lines, selecting outstanding candidates, and intermating individual plants to generate novel and superior allelic combinations. While significant efforts have been made to integrate digital tools and workflows into the first two steps of this cycle, readily-available digital tools for tracking crosses in plant breeding are non-existent or have become obsolete (Rife and Poland, 2014; Perez, 2017; Courtney et al., 2018).

Organized intermating of individuals is the backbone of any plant breeding program and is necessary to develop diversity for subsequent selection or to create populations to dissect the genetic control of phenotypes. However, most major breeding programs still rely on paper, handwritten tags to track crosses which can lead to errors and inaccuracies. Mistakes in pedigrees are not uncommon and unexpected gene frequencies or progeny are often attributed to pedigree errors (Fang et al., 2011; Endelman et al., 2017; Bhatta et al., 2018; Montanari et al., 2020). While modern genotyping systems can assist in the identification and correction of pedigree inaccuracies, many plant breeding programs do not have access to or have not introduced the necessary workflows to facilitate these corrections (Endelman et al., 2017; Grattapaglia et al., 2018; Fradgley et al., 2019).

# IMPLEMENTATION

We have designed and developed Intercross, an Android application, that aims to be a simple addition to plant breeding workflows. Intercross is freely available and open source, allowing breeding programs to make code contributions or further customize it for their own needs.

At its most basic level, Intercross works to associate a male ID, female ID, and cross ID while simultaneously capturing a timestamp and the name of the person making the cross (Figure 1a). The cross ID can be supplied by the breeder using pre-printed labels, generated as a unique universal ID, or designed to match a custom pattern (Figure 1b).

Based on the parental combination, Intercross determines if the cross is biparental, self-pollinated, open-pollinated, or a polycross.

Intercross can be further extended to capture the entire workflow of generating a new seedlot or family. After the cross has been created, the cross can optionally be advanced through a workflow that covers the entire lifecycle of that cross. This enables the user to track additional information including the number of flowers used for the cross, the number of harvested items (e.g. fruits, pods, spikes, etc.), and the ultimate number of seeds created from a specific crossing event (Figure 1c).

The integration of barcodes significantly minimizes error when utilizing complex sample identifiers. Many facets of plant breeding workflows can be optimized by using barcodes. Intercross has built-in barcode scanning support and allows users to search for specific cross IDs via barcode to rapidly navigate to an individual cross page.

Breeders making crosses regularly identify specific plants in crossing blocks by their common cultivar or family name and not the barcoded ID. To bridge these two different naming systems, Intercross allows users to import a list of parents that includes the barcode ID, the sex of the parent, and the common name of the parent into the app (Figure 2a, Table S1). When these barcode IDs are used to make crosses, they are visually replaced with the parent names (Figure 2b). Users can actively manage this parent list in the app by adding females, males, or pollen groups (e.g. for polycrosses) with UUIDs or custom barcode IDs.

To facilitate tracking crosses in programs creating polycrosses, Intercross allows users to group individuals, generate a unique identifier, and print a barcoded label. When exporting polycrosses, the male parent is replaced with a list of all males that were used to make the pollen group.

Intercross can be paired with Zebra Bluetooth printers (Zebra Technologies Corporation, Lincolnshire, Illinois, USA) to print both individual parent and cross labels. Labels can be further formatted to a users' specification by importing a custom ZPL template file.

Crossing blocks are generally designed each season with an idea of which and how many specific crosses should be made. Creating additional crosses beyond what is needed for any given parental combination is often seen as an inefficient use of resources. Intercross allows breeders to import a wishlist that is used to track progress toward a desired production threshold (Table S2). Breeders can track their progress toward the number of targeted crosses via two different interfaces: a simple list with the different combinations and a grid with specific parental combinations highlighted.

Collected data can be exported to a CSV (comma separated value) file in the internal device memory with each row including the cross ID, both parent IDs, a timestamp, the person who made the cross, the type of cross, and any additional metadata that was collected about the crossing event. Additional technical details of Intercross are described in Courtney and Neilsen (2019).

In addition to the features above, settings have been added to allow a blank male ID (e.g. for open-pollinated crosses), change the order that the parent boxes are listed on the cross creation page, and toggle audio notifications to alert the user when a cross is successfully created.

We have worked with breeders across a diverse group of crops to identify a suite of features for Intercross that covers the majority of needs to manage and track crosses. Future updates could expand the app to track emasculation events and allow fine-grained time tracking. We appreciate that with real-world use, features and workflow improvements that have not been considered will be identified and requested based on the disparate methods in use among breeding programs, and we look forward to working with the plant breeding community to further improve Intercross.

Ultimately, Intercross will fully integrate into breeding database workflows by implementing the Breeding API (BrAPI) which will simplify digital breeding workflows (Selby et al., 2019).

**PERSPECTIVE**

Tracking crossing events and parental contributions within plant breeding and genetics programs is necessary to maintain a successful breeding system. Pedigree information is the backbone of breeding and is often utilized for research projects wholly unrelated to primary breeding goals. Ensuring that the data being collected is accurate is in the interest of all members of the breeding community. Utilizing digital tools for data management to minimize errors and maximize efficiency should be a priority for all scientists. Intercross creates the framework necessary to capture and manage crossing-related data for plant breeders and geneticists. By bridging the gap between parent-progeny digital collection, Intercross further promotes the concept of a digital breeding ecosystem to maximize the efforts of breeders around the world and contribute toward the improved varieties necessary for global food security.

**DATA AVAILABILITY**

Source code for Intercross can be found on GitHub:

https://github.com/PhenoApps/Intercross

Intercross can be found in the Google Play Store:

https://play.google.com/store/apps/details?id=org.phenoapps.intercross

**CONFLICT OF INTEREST**

None declared.

**ACKNOWLEDGEMENTS**

The authors wish to express gratitude to the breeders and technicians at The International Potato Center and the International Institute of Crop Agriculture who contributed ideas and willingly tested early versions of Intercross, including Elisa Salas, Jolien Swanckaert, Prasad Peteti, Ismail Rabbi, and Afolabi Agbona. We also appreciate the valuable suggestions and critiques provided by Lukas Mueller, Bryan Ellerbrock, and Mary Guttieri.

**FUNDING**

The development of Prospector was supported by the National Science Foundation under Grant No. (1543958). This app is made possible by the support of the American People provided to the Feed the Future Innovation Lab for Crop Improvement through the United States Agency for International Development (USAID). The contents are the sole responsibility of the authors and do not necessarily reflect the views of USAID or the United States Government. Program activities are funded by the United States Agency for International Development (USAID) under Cooperative Agreement No. 7200AA-19LE-00005.

**FIGURES AND TABLES**

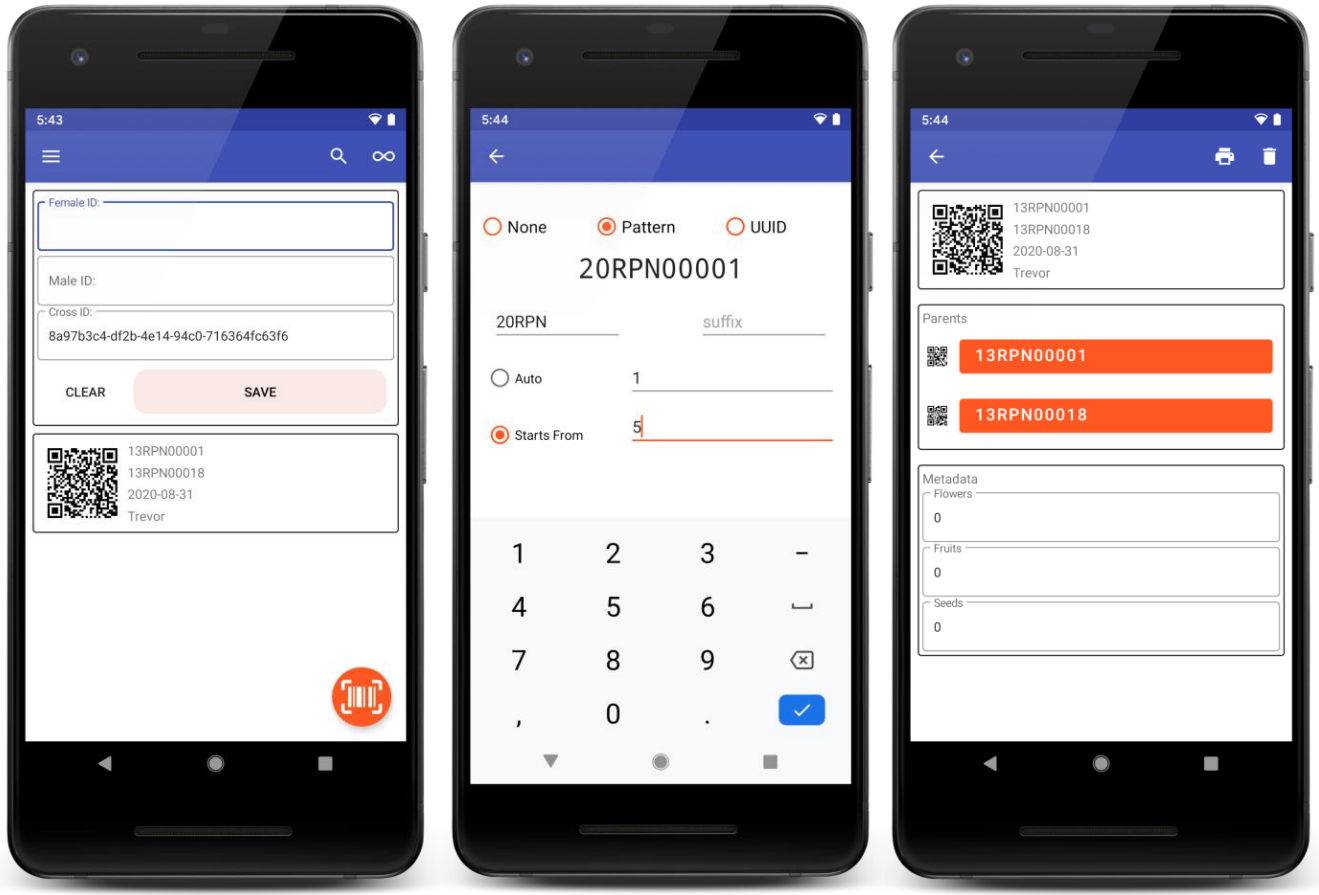

*Figure 1. a) The primary interface for Intercross. Crosses are created by entering or scanning a Female ID, a Male ID, and a Cross ID and pressing Save. The created cross and QR code of its ID is shown in list below the data entry area. b) Cross IDs can be input by the user, constructed to match a specific pattern (shown), or programmatically generated to be a universal unique ID (UUID). When creating a pattern, users can specify a prefix (e.g. 20RPN), a suffix, the first number for the cross (e.g. 1), and the length of the padding value (e.g. 5). c) A cross that has been created shows the generated label (top), both parents (middle), and optional metadata entry area (bottom).*

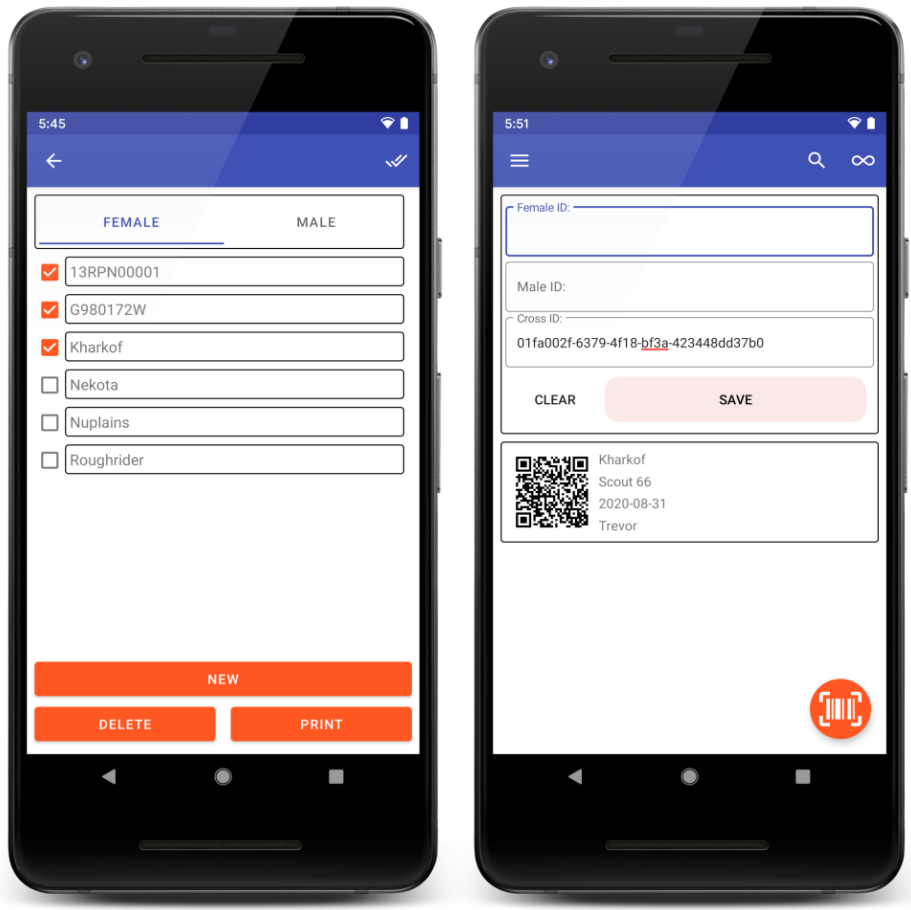

*Figure 2. a) The primary interface for the Parents' manager. Parents can be manually created or imported from a premade list. Barcoded labels can be printed for selected parents. b) When the ID for a parent is used in a cross, the parent name is displayed instead of the ID (e.g. Kharkof and Scout 66 instead of 15RPN00001 and 15RPN00018.*

Table S1. Parent import for Intercross

| codeId | sex | name |
|---|---|---|
| 15RPN00001 | 0 | Kharkof |
| 15RPN00018 | 1 | Scout 66 |
| 15RPN00002 | 0 | Roughrider |
| 15RPN00019 | 1 | Kharkof |
| 15RPN00003 | 0 | Nuplains |
| 15RPN00020 | 1 | TAM-107 |
| 15RPN00004 | 0 | Nekota |
| 15RPN00021 | 1 | SD05W140 |
| 15RPN00005 | 0 | G980172W |
| 15RPN00022 | 1 | SD02804-1 |

Table S2. Wishlist file for Intercross

| femaleDbId | maleDbId | femaleName | maleName | wishType | wishMin | wishMax |
|---|---|---|---|---|---|---|
| 13RPN00001 | 13RPN00018 | Kharkof | Scout 66 | cross | 5 | 10 |
| 13RPN00001 | 13RPN00019 | Kharkof | Kharkof | cross | 1 | 5 |
| 13RPN00001 | 13RPN00020 | Kharkof | TAM-107 | cross | 2 | 2 |
| 13RPN00002 | 13RPN00019 | Roughrider | Kharkof | cross | 1 | 5 |
| 13RPN00003 | 13RPN00020 | Nuplains | TAM-107 | cross | 1 | 3 |
| 13RPN00004 | 13RPN00021 | Nekota | SD05W140 | cross | 1 | 100 |
| 13RPN00005 | 13RPN00022 | G980172W | SD02804-1 | cross | 1 | 7 |
| 13RPN00006 | 13RPN00023 | G980723 | SD05004 | cross | 1 | 5 |
| 13RPN00007 | 13RPN00024 | G980926 W | SD05118 | cross | 1 | 5 |
| 13RPN00008 | 13RPN00025 | NW97S412-1 | SD03164-2 | cross | 1 | 5 |
| 13RPN00009 | 13RPN00026 | Kharkof | SD05118 | cross | 1 | 5 |
| 13RPN00010 | 13RPN00027 | Harding | SD05210 | fruit | 3 | 9 |
| 13RPN00011 | 13RPN00028 | Nuplains | SD05W030 | fruit | 3 | 9 |
| 13RPN00012 | 13RPN00029 | Nekouta | Antelope | flower | 3 | 9 |
| 13RPN00013 | 13RPN00030 | Trego | Wesley | flower | 3 | 9 |
| 13RPN00014 | 13RPN00031 | T140 | Jerry | seed | 3 | 9 |
| 13RPN00015 | 13RPN00032 | T141 | NX05M4180-6 | seed | 3 | 9 |
| 13RPN00016 | 13RPN00033 | T138 | NW04Y2188 | fruit | 3 | 9 |
| 13RPN00017 | 13RPN00034 | SD00151-7 | HV9W04-1186W | fruit | 3 | 9 |